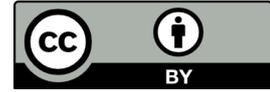

# Bipolar single-molecule electroluminescence and electrofluorochromism

Tzu-Chao Hung,[1] Roberto Robles,[2] Brian Kiraly,[1] Julian H. Strik,[1] Bram A. Rutten,[1] Alexander A. Khajetoorians,[1] Nicolas Lorente,[2,3] and Daniel Wegner[1]*

1. Institute for Molecules and Materials, Radboud University, Nijmegen, The Netherlands
2. Centro de Física de Materiales, CFM/MPC (CSIC-UPV/EHU), Paseo de Manuel de Lardizabal 5, 20018 Donostia-San Sebastián, Spain
3. Donostia International Physics Center (DIPC), Paseo de Manuel de Lardizabal 4, 20018 Donostia-San Sebastián, Spain
* Corresponding author: d.wegner@science.ru.nl

Date: June 13, 2023

**ABSTRACT**

Understanding the fundamental mechanisms of optoelectronic excitation and relaxation pathways on the single-molecule level has only recently been started by combining scanning tunneling microscopy (STM) and spectroscopy (STS) with STM-induced luminescence (STML). In this paper, we investigate cationic and anionic fluorescence of individual zinc phthalocyanine (ZnPc) molecules adsorbed on ultrathin NaCl films on Ag(111) by using STML. They depend on the tip-sample bias polarity and appear at threshold voltages that are correlated with the onset energies of particular molecular orbitals, as identified by STS. We also find that the fluorescence is caused by a single electron tunneling process. Comparing with results from density functional theory calculations, we propose an alternative many-body picture to describe the charging and electroluminescence mechanism. Our study provides aspects toward well-defined voltage selectivity of bipolar electrofluorochromism, as well as fundamental insights regarding the role of transiently charged states of emitter molecules within OLED devices.

**POPULAR SUMMARY**

In organic light emitting diodes (OLEDs), a current through the device leads to generation of light, referred to as electroluminescence. OLEDs are very promising for highly energy-efficient light sources. Surprisingly, the underlying fundamental processes that excite the emitter molecules and eventually lead to light emission are not trivial and often unclear. We combined the atomic-scale spatial resolution of scanning tunneling microscopy with light detection in order to simultaneously study both the molecular electronic structure and the fluorescence emitted from individual zinc phthalocyanine (ZnPc) molecules embedded in the tunnel junction. Through a systematic approach, we were able to precisely identify all involved processes of charging, excitation, relaxation, and eventually light emission, as consistently summarized in a many-body energy diagram.

Our study revealed that in most cases, light emission of ZnPc in the electroluminescent tunnel junction is initiated by charging the molecule, into an either cationic or anionic state, depending on the polarity of the applied voltage. Consequentially, we discovered that the molecule can exhibit bipolar electrofluorochromism, i.e., the emission color can be changed by reversing the voltage. Threshold voltages of emission lines can be rationalized by tunneling into or out of particular molecular orbitals. Supported by time-dependent density function theory, we propose that Auger-like intramolecular electronic relaxation processes have to be considered to understand fluorescence from charged ZnPc. Our systematic approach can serve as blueprint for future studies of electroluminescent molecules and shows that transiently charged states of emitter molecules should also be considered in the design optimization of OLEDs.

## I. INTRODUCTION

Molecular optoelectronic excitation and relaxation pathways involving intermediate charged (i.e. radical) states are not only important for many chemical reactions, but they also play a crucial role in organic electronics devices, where electrons and holes are injected to activate a molecular function, e.g. in sensors [1] or organic light-emitting diodes (OLEDs) [2]. For the latter, a fundamental understanding of the interplay between charge transfer into and exciton formation within chromophores may provide important insight toward designing more efficient OLEDs, e.g. based on triplet emitters [3]. On the single-molecule level, opto-electronic properties of individual molecules can be probed in nanojunctions [4,5], but they lack information on how the molecule is embedded in its local environment. Scanning tunneling microscopy (STM) not only provides the necessary spatial resolution to resolve the adsorption configuration of a single molecule on a surface, but it grants control of the local environment around the molecule with atomic-scale precision in conjunction with atomic manipulation techniques. Combining this with the detection of light emitted at the tunnel junction leads to measuring the STM-induced luminescence (STML) with sub-molecular resolution [6-13]. This allows for fundamentally understanding how molecular electronic excitations and radiative decay occur in a tunneling transport device. Initial studies debated whether the excitation occurs via pure charge injection [8,12] or by energy transfer from plasmons in the tip-sample nanocavity [9], but the consensus is that both mechanisms can occur simultaneously and couple coherently [10,13-15].

Adding to this scientific discussion, recent STML studies discovered that fluorescence from the radical cationic state of zinc phthalocyanine (ZnPc) [16,17] and free-base phthalocyanine ($H_2Pc$) [18] can be observed if a sufficiently large negative sample bias voltage is applied in the tunnel junction. This is possible because the molecule, which is decoupled from the metal substrate by ultrathin NaCl films, can stabilize the charged $[ZnPc]^+$ state sufficiently long enough to allow formation and radiative decay of an exciton before the molecule goes back into the neutral state. Two very different mechanisms of exciton formation have been proposed, both requiring a two-electron tunneling process. One proposed scenario is based on plasmon-molecule interactions [16,18], whereas the other is based on pure charge carrier injections [17]. Both proposals, however, may not resolve the observed threshold voltages required to produce the cationic emission.

In this paper, we perform a systematic combined STM, STS and STML study on ZnPc molecules adsorbed on ultrathin NaCl films on Ag(111). We correlate electronic and optical properties and quantitatively identify and characterize the thresholds at which certain emission features are observed. Our measurements show that ZnPc displays bipolar electrofluorochromism, i.e., it can be transiently charged both positively ([ZnPc]$^+$) and negatively ([ZnPc]$^-$), depending on the polarity of the sample bias voltage, as seen by characteristic charged-state emission in STML. We find that threshold voltages are connected to accessibility of particular molecular orbitals (MOs) in tunneling transport. Supported by density functional theory (DFT) and time-dependent DFT (TDDFT) calculations, we propose a mechanism in which Auger-like intramolecular electronic relaxation processes are key to understand the fluorescence from charged ZnPc, rationalizing the required threshold voltages. Furthermore, we present a many-body diagram that describes in a consistent picture not only the molecular excitation and relaxation pathways of neutral, cationic, and anionic ZnPc in our study, but also observations previously reported on other substrates [16-18].

## II. Experimental details

The experiments were carried out in a commercial Omicron ultra-high vacuum (UHV) low-temperature STM system operated at $T$ = 4.5 K with a base pressure below 1×10$^{-10}$ mbar [13]. The silver bulk tip was electrochemically etched in a mixture of perchloric acid and methanol (ratio 1:4) [19,20], and further treated in UHV by field emission and controlled indentation on a Ag(111) surface. The Ag(111) single crystal (MaTeck) was cleaned by multiple cycles of sputtering and annealing followed by NaCl deposition while the Ag(111) surface was kept at room temperature. Single ZnPc molecules were deposited on the surface held at $T$ < 6 K inside the STM. All STML spectra were acquired using a 150 grooves/mm grating. Further detail of the optical setup can be found in a previous publication [13].

## III. RESULTS

Figure 1(a) shows a constant-current STM image of ZnPc molecules adsorbed on terraces of two and three monolayers (ML) NaCl supported by Ag(111). On 2 ML, the image shows an individual ZnPc, displaying 16 apparent lobes, due to the rapid shuttling motion [13]. In addition, a ZnPc dimer is seen on both the 2 and 3 ML NaCl terrace, where 8 lobes are visible for each ZnPc due to the interlocking of the molecules that suppresses the rapid shuttling. These dimers usually cannot be found in as-grown samples, and we have assembled them by means of STM manipulation [8,13]. We note that this is necessary for the acquisition of STML spectra at positive sample bias, as the adsorption position of monomers is not sufficiently stable under these tunneling conditions. Alternatively, ZnPc monomers can be also stabilized by anchoring to a clean NaCl step edge (see Appendix A and C) [13]. All these experiments are comparable, because the electronic structure of monomers and dimers on NaCl layers with equal thickness is identical, as reveled by STS.

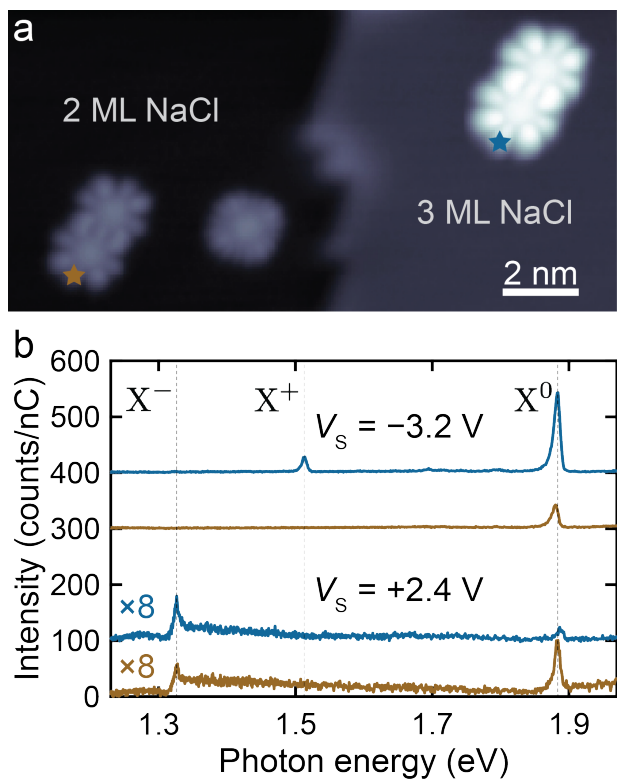

**Figure 1. ZnPc fluorescence on 2 vs. 3 ML NaCl/Ag(111).** (a) Constant-current STM image of ZnPc molecules adsorbed on 2 and 3 ML NaCl/Ag(111) ($V_S = -2.5$ V, $I_t = 10$ pA). In order to acquire STML spectra at positive sample bias, the ZnPc molecules were assembled into dimers by STM manipulation. (b) STML spectra of the ZnPc dimers on 2 ML and 3 ML NaCl/Ag(111) at opposite sample bias polarities ($V_S$ is indicated in the figure, $I_t = 100$ pA, acquisition time $t = 120$ s) measured at the positions marked in (a). All spectra were offset for clarity.

Figure 1(b) presents STML spectra of both ZnPc dimers, acquired with the tip positioned as marked in Fig. 1(a). With an applied sample bias voltage of $V_S = -3.2$ V (top two spectra), the STML spectra reveal a main peak at a photon energy of 1.880 eV on 2 ML NaCl (brown curve), and at 1.883 eV on 3 ML

NaCl (blue curve). This peak was previously identified as the Q(0,0) transition of the ZnPc molecule, which is slightly red-shifted in the dimer due to dipole-dipole coupling [8,9,21]. Here, we refer to it as $X^0$ in Fig. 1(b) and the following discussion. We note that the $X^0$ peak intensity strongly increases on 3 ML NaCl compared to the case on 2 ML NaCl. For the dimer on 3 ML NaCl, an additional peak is found at a photon energy of 1.513 eV, which is not visible for the dimer on 2 ML NaCl. This peak was previously identified as the emission of a positively charged (i.e. cationic) $[ZnPc]^+$ molecule [16], referred to as $X^+$ in Fig. 1(b) and the following discussion. In agreement with previous studies, no red-shift of the $X^+$ peak is found in the dimer, implying that only the molecule under the tip is transiently charged by electron tunneling (see also appendix A) [22].

At positive applied sample bias ($V_S$ = +2.4 V), we observe a previously unreported fluorescence peak that we assign to anionic emission. The STML intensity is about a factor of ten smaller compared to the STML acquired at negative sample bias. Despite of the low STML intensity, the ZnPc dimers show the $X^0$ emission at 1.883 eV and 1.886 eV on 2 and 3 ML NaCl (brown and blue curves in lower part of Fig. 1(b)), respectively. An additional peak is observed at 1.326 eV on both 2 and 3 ML NaCl. The overall observations of this feature are reminiscent of the $X^+$ peak at negative bias, i.e., no red-shift occurs between monomer and dimer STML, and the peak vanishes when the tip is placed next to the molecule and only tunneling into the substrate occurs (see Appendix A). By comparison with the literature, the energy of this feature is in excellent agreement with experimental and calculated optical spectra of the negatively charged (i.e. anionic) $[ZnPc]^-$ molecule [23-25]. Therefore, we refer to this peak as $X^-$ in Fig. 1(b) and the following discussion. Interestingly, the intensity of the $X^-$ emission increases from 2 to 3 ML NaCl, whereas that of $X^0$ decreases. As we can observe all three STML peaks $X^0$, $X^+$, and $X^-$ on 3 ML NaCl/Ag(111), we focus our further discussion on ZnPc dimers on that NaCl thickness.

To further elucidate under which conditions fluorescence of the neutral, cationic and anionic molecule occurs within the STM tunnel junction, we performed sample bias-dependent STML measurements. Note that, while the experiment is a double tunneling barrier, it was shown that the applied voltage almost exclusively drops in the vacuum barrier between tip and molecule, i.e., the voltage drop across the NaCl barrier is negligibly small [9,13,14]. STML spectra for $V_S$ < 0 with various magnitudes are presented in Fig. 2(a). Again, the $X^0$ peak at 1.883 eV is the main feature. It becomes visible at $V_S \lesssim$

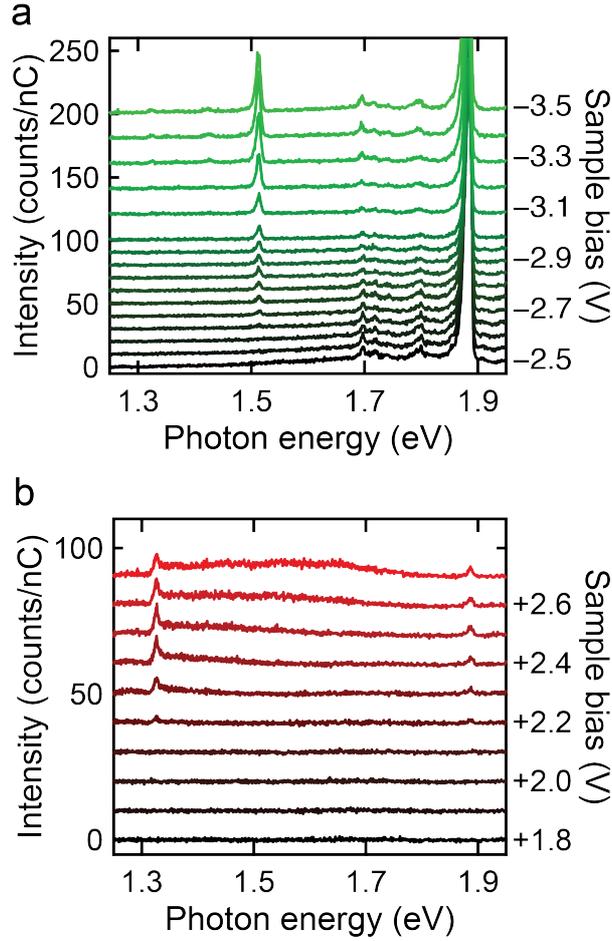

**Figure 2. Bias-dependent STML spectra of ZnPc dimer on 3 ML NaCl/Ag(111).** (a) Negative bias-dependent STML spectra taken with decreasing sample bias. The emission of $X^+$ is observed when the $V_S \leq -2.65$ V. (b) Positive bias-dependent STML spectra taken with increasing sample bias. The emissions of $X^0$ and $X^-$ are both observed when the $V_S \geq +2.20$ V. All STML spectra in (a) and (b) are raw data and were taken in constant-current mode at the tip position as indicated in Fig. 1(a) ($I_t$ = 100 pA, $t$ = 120 s).

−1.9 V (see Appendix B) but has a strongly increased intensity for $V_S \leq -2.2$ V. At lower photon energies down to about 1.7 eV, vibronic "loss" peaks can be seen whenever the $X^0$ intensity is sufficiently strong [9], but without displaying any dependence on $V_S$. The $X^+$ peak appears for sample biases $V_S \leq -2.65$ V and its intensity increases with further decreasing $V_S$. For different tips used in our experiments, we observed slight variations of the threshold voltage of $X^+$, ranging between −2.65 V and −2.90 V (see Appendix B for more details), in accordance with previous observations [16,17]. STML spectra for $V_S >$ 0 are shown in Fig. 2(b). Here, the situation turns out to be much simpler. Both the $X^0$ and $X^-$ peaks are found at $V_S \geq +2.2$ V. Especially, we can rule out that emission starts at $V_S \approx +1.9$ V for the $X^0$ peak.

A valuable piece of information to determine the excitation and relaxation mechanisms leading to $X^0$, $X^+$ and $X^-$ emissions, is a power-law analysis of the STML intensity $\Im_{\text{photon}}$ vs. tunnel current $I_t$, which are related via $\Im_{\text{photon}} \propto I_t^\alpha$ [16,18,26,27]. The power $\alpha$ indicates how many tunneling electrons are involved in the STML process, i.e., a single-electron process for $\alpha = 1$ or a two-electron process for $\alpha = 2$. However, we find that for negative bias voltages the power laws for both $X^0$ and $X^+$ are strongly tip-dependent, varying in the range of $1.09 \leq \alpha(X^0) \leq 1.49$ and $1.28 \leq \alpha(X^+) \leq 2.00$, respectively (see Appendix B). We propose that the tip dependence at negative bias is a consequence of the contribution of nanocavity plasmons in the emission process of both $X^0$ and $X^+$ [21,28-30]. On the other hand, at positive polarity the power analysis gives a very clear picture: we find $0.81 \leq \alpha(X^0) \leq 0.94$ and $1.06 \leq \alpha(X^-) \leq 1.09$. This much narrower range means that the power law is essentially tip-independent, and the values strongly suggest a one-electron process for both emissions. At this point it becomes clear that, at least for the $X^-$ and $X^0$ at positive bias, an alternative mechanism is required in which a single tunneling electron both charges the ZnPc and leads to exciton formation (i.e. the creation of an electron-hole pair) within the molecule.

Next, we rationalize the observed threshold voltages by comparing them to onset energies of MOs, as found from STS. In Fig. 3(a), the differential conductance, $dI_t/dV_S$, of ZnPc adsorbed on 3 ML NaCl/Ag(111) is shown with respect to the sample bias. In rough approximation, $dI_t/dV_S$ is proportional to the local density of states (LDOS) and the Fermi level ($E_F$) is located at $V_S = 0$ V. The onset of the first negative ion resonance (NIR) [31] is found at $V_S = 0.6$ V, and it has a peak at $V_S = 0.8$ V. This state can be interpreted as the superposition of LDOS from the two degenerate lowest unoccupied molecular orbital (LUMO) and LUMO+1 [12,13,16]. A slight shift of the onset of the first NIR is observed between the STS acquired at the center (blue curve in Fig. 3(a)) and at the lobe (red curve in Fig. 3(a)). This can be attributed to the different tip-sample displacement-induced voltage drop in the double tunneling barrier [32,33]. The STS feature is known to be broadened due to electron-phonon coupling [32], and it was shown that the onset energy is a better measure for the position of the LUMO [14]. The orbital assignment to the LUMO/+1 is confirmed by acquiring a spatial map of the $dI_t/dV_S$ peak intensity, shown in Fig. 3(c) [13,16]. The onset of the second NIR is found at $V_S = +2.1$ V, which interestingly corresponds to the bias threshold of both $X^0$ and $X^-$ emissions at positive sample bias. A similar relationship between the STML threshold and the STS onset of the second NIR is also found on the ZnPc dimer adsorbed on 2 ML NaCl/Ag(111) (see Appendix C). The corresponding $dI_t/dV_S$ map (Fig. 3(d)) shows a more cross-like distribution, different from the LUMO/+1 state. This becomes more obvious when acquiring maps on single molecules (see Fig. 11 in Appendix D). On the one hand, we do not observe any additional LDOS peak between the LUMO/+1 and this state, hence it may be assigned to the LUMO+2. On the other hand, comparison of the orbital distribution with our DFT calculations suggests that this may be the LUMO+3, and we cannot rule out that the LUMO+2 may be obscured in STS (see Appendix D for a detailed discussion). Therefore, we refrain from a definitive assign and refer to the second NIR as the LUMO+$x$ (where x ≥ 2).

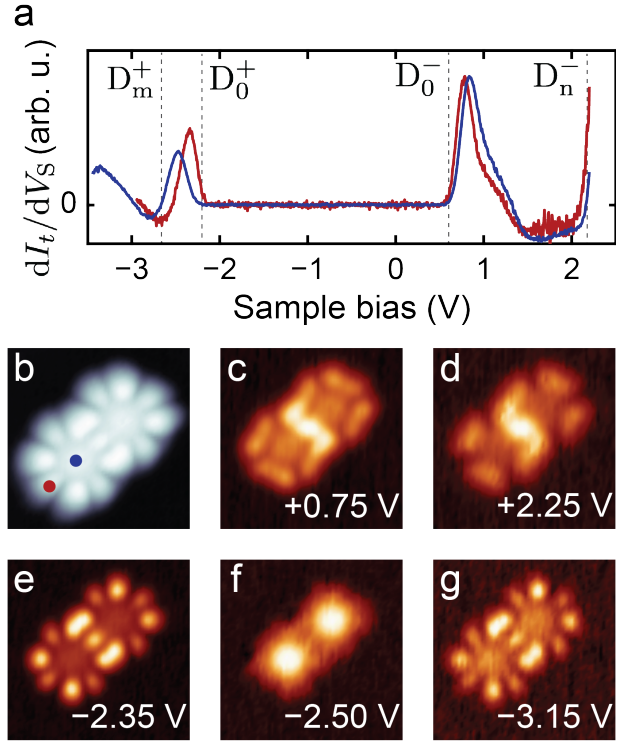

**Figure 3. Characterization of the molecular orbitals.** (a) STS of the ZnPc dimer at the center (blue) and lobe (red) of the molecule (tip position indicated by dots in (b), feedback loop opened at $V_S = -3.0$ V, $I_t = 50$ pA (blue) and $I_t = 100$ pA (red), respectively). (b) Constant-current STM image of ZnPc dimer adsorbed on 3 ML NaCl/Ag(111) ($V_S = -2.5$ V, $I_t = 10$ pA). (c)-(g) The constant-height differential conductance maps show (c) the LUMO/LUMO+1, (d) LUMO+x, (e) HOMO, (f) HOMO−1 and (g) HOMO−x ($V_S$ as stated in the figures, lock-in parameters $V_{rms} = 8$ mV at $f_{mod} = 819$ Hz, feedback loop was opened at the center of molecule at $I_t = 50$ pA, and sample bias (c) $V_S = +0.8$ V, (d) $V_S = +2.3$ V, (e,f) $V_S = -2.5$ V, and (g) $V_S = -3.3$ V).

At negative bias, the onset of the first positive ion resonance (PIR) is found at $V_S$ = −2.15 V when the tip is parked on the lobe of the molecule (red curve in Fig. 3(a)) and the onset of the second PIR is found at $V_S$ = −2.25 V when the tip is parked at the center of the molecule (blue curve in Fig. 3(a)). With the help of the spatial maps (Fig. 3(e,f)) as well as DFT calculations (Appendix D), the first and the second PIR are assigned to the highest occupied molecular orbital (HOMO) and the HOMO−1, respectively, the latter being mainly located at the Zn atom and exhibiting significant *d*-orbital character. The $dI_t/dV_S$ spectra do not show a sharp onset of a PIR at the threshold of the $X^+$ emission. However, a region of negative differential resistance (NDR) is visible in this bias region, which may artificially shift a MO onset, making it difficult to pinpoint. Nevertheless, the $dI_t/dV_S$ intensity clearly increases with further decreasing bias, revealing LDOS contributions from one or more occupied orbitals below the HOMO–1. To further investigate if a new MO appears around the threshold of $X^+$, we perform $dI_t/dV_S$ mapping with increment bias steps (see Appendix D). We indeed find a transition of LDOS distribution from the HOMO–1 to another MO within the bias range of the $X^+$ threshold. Figure 3(g) shows a spatial map at $V_S$ = –3.15 V, reminiscent of the maps taken within the NDR region. From the comparison with the calculations, it is not clear which MO is probed, presumably due to the broad Gaussian line shapes caused by electron-phonon coupling [14,31,32], but it must be either the HOMO–2 or an even lower lying MO, and we refer to this as the HOMO–*x*.

## IV. DISCUSSION

Our results from combining STML and STS allow us to propose an alternative excitation mechanism that can explain all observations in a consistent picture based on intramolecular relaxation processes. Key to our interpretation is the requirement that all observed emission peaks stem from a single tunneling event into (or out of) a particular MO (with the only exception being the $X^0$ emission in the range –2.15 V < $V_S$ < 1.9 V, where plasmon-molecule coupling leads to exciton formation [9,16]). Supported by TDDFT calculations (see Appendix E for details), we propose a many-body energy diagram that can explain all observations and whose levels can be connected to particular orbital occupations in the single-particle picture. This energy diagram is shown in Fig. 4 and will be discussed in the following.

We first focus on explaining the diagram for processes occurring at $V_S > 0$, where the molecule is either in a neutral (i.e. singlet S) or an anionic (i.e. doublet $D^-$) state. The molecule initially is in the singlet ground state ($S_0$). From the STS data, tunneling into the LUMO/+1 occurs for $V_S \geq 0.6$ V, leaving the molecule in the anionic doublet ground state $D_0^-$ (blue solid arrow (1) in Fig. 4). As no $X^0$ emission is observed for $V_S < 2.1$ V, the $D_0^-$ level must be below the excited singlet state $S_1$ of the neutral molecule, such that the molecule can only go back to $S_0$ via a discharging process (dashed gray arrow (i)) without any light emission. At $V_S \geq 2.1$ V, resonant electron tunneling into the second NIR (LUMO+$x$) occurs. Now, the molecule is excited into a higher anionic doublet state $D_n^-$ (blue solid arrow (2) in Fig. 4). From there, the molecule can electronically relax into the lowest optically active excited doublet state $D_1^-$ (dashed black arrow (a) in Fig. 4), in which an exciton has formed in the charged molecule. We note that the $D_1^-$ state was also referred to as a trion state [17]. In a simplified single-particle picture, the electron relaxes from the LUMO+$x$ to one of the LUMO/+1 levels, thereby transferring the excess energy to an electron in the HOMO, which consequently also gets excited to the LUMO/+1 and leaves a hole in the HOMO. This Auger-like intramolecular relaxation process causes the exciton formation in the charged molecule, without the need for any additional tunneling process. Our TDDFT calculations (see Appendix E) support this picture and confirm that the total energy of the anionic ZnPc decreases, i.e. $E(D_n^-) > E(D_1^-)$, such that this relaxation is energetically feasible.

Once in the $D_1^-$ state, the molecule can either remain charged and relax into the $D_0^-$ ground state, or it discharges back into the neutral state. In case of the former, the relaxation corresponds to a recombination of the electron-hole pair from the LUMO/+1 and HOMO, and the excess energy is emitted as a $X^-$ photon. In case of the latter, one of the electrons in the LUMO/+1 tunnels to the Ag(111) substrate. As there is still an electron-hole pair present, this discharging pathway goes to the excited singlet state $S_1$ (dashed gray arrow (ii) in Fig. 4), from which the molecule can radiatively decay into the singlet ground state, $S_0$, emitting a $X^0$ photon. In order to allow for these two alternative pathways to be possible, the $D_1^-$ state must be above the $S_1$ state in our many-body energy diagram. A consequence of this diagram is that $X^-$ and $X^0$ emissions compete. This is confirmed in the STML spectra at positive bias shown in Fig. 1(b): on 2 ML NaCl, the $X^0$ emission is more intense than the $X^-$ emission, whereas on 3 ML NaCl the situation is reversed. The trend is also confirming the picture, as a thinner NaCl film reduces the tunnel barrier to the substrate, favoring a discharging event and hence $X^0$ emission, whereas the thicker insulator will stabilize the anionic state and favor $X^-$ emission. We note that the electron configuration corresponding to the $D_1^-$ state cannot be directly accessed by a single-electron tunneling event. In principle, it would be possible that a sufficiently large voltage below the LUMO+*x* onset could excite the charged molecule from $D_0^-$ to $D_1^-$ via energy transfer from the plasmon resonance. However, we can rule out this excitation pathway, as it would require a two-electron process as well as observation of $X^-$ emission at sample voltages below the LUMO+*x* onset, both of which we do not observe.

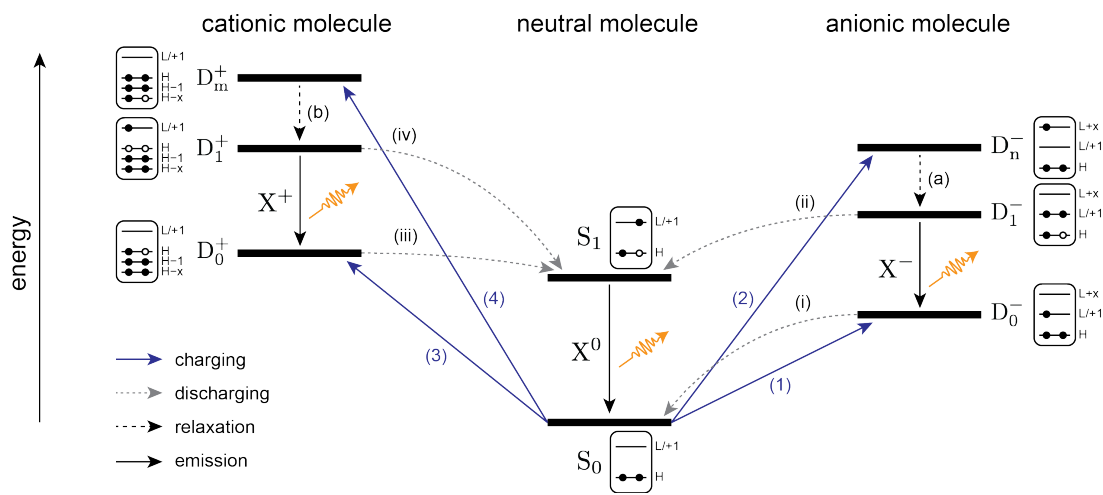

**Figure 4. Schematic many-body energy diagram for exciton formation of ZnPc adsorbed on 3 ML NaCl/Ag(111).** The light emission of $X^+$ ($X^-$) originates from the transition of excited doublet state $D_1^+$ ($D_1^-$) to doublet ground state $D_0^+$ ($D_0^-$). To reach the $D_1^+$ ($D_1^-$) state, tunneling into/out of higher energy MOs is required which leads to excitation into $D_m^+$ ($D_n^-$). Relaxation to $D_1^+$ ($D_1^-$) occurs via Auger-like intramolecular transitions. The excited singlet state (S$_1$) can be reached by discharging from the $D_1^\pm$ and/or from $D_0^+$, but not from the lower-lying $D_0^-$. From there, $X^0$ emission occurs by radiatively decaying into singlet ground state (S$_0$). Note that the energy axis is not to scale, i.e., the positions of the levels are only qualitative in nature.

Next, we discuss the energy diagram for processes occurring at $V_S < 0$, where the molecule is either in a neutral singlet or a cationic doublet ($D^+$) state. Again, the molecule is initially in the $S_0$ state. Plasmon-mediated excitation to $S_1$ (not shown in Fig. 4) is already possible at voltages $V_S \leq -1.9$ V, below the onset of the HOMO [9]. However, this process is overall weak, leading to relatively small $X^0$ emission intensities (see Fig. 7 in Appendix B) [7,27]. The situation changes drastically as soon as tunneling out of the HOMO is possible at $V_S \leq -2.15$ V. During this charging process (solid blue arrow (3) in Fig. 4), the molecule goes into the cationic doublet ground state $D_0^+$. As shown previously, this state is higher in energy than the $S_1$ state [14]. Therefore, the molecule can be discharged into $S_1$ (dashed gray arrow (iii) in Fig. 4) by an electron tunneling from the Ag(111) substrate into the LUMO/+1, and from there radiative decay to the $S_0$ state is possible via $X^0$ emission. We note that this process is plasmon-enhanced, leading to a tip-dependent non-integer power law (Fig. 8 in Appendix B) as well as a Fano line shape of the $X^0$ STML spectrum [13].

For the observation of $X^+$ emission, tunneling out of a deeper lying occupied orbital (HOMO–$x$) is required. We note that experimentally it is not clear which orbital is involved, but the TDDFT calculations show that $x = 2$ is already sufficient. Therefore, in the following discussion we will assume tunneling out of the HOMO–2, noting that all processes would also be possible for $x > 2$. Tunneling out of this orbital (solid blue arrow (4) in Fig. 4) will leave behind a hole in the HOMO–2, leading to a highly excited cationic doublet state $D_m^+$. From there, the molecule can electronically relax into the lowest excited doublet state $D_1^+$ (dashed black arrow (b) in Fig. 4), in which an exciton has formed in the charged molecule. Again, the interpretation in a simplified picture of orbital occupation is that of an intramolecular Auger-like transition where an electron in the HOMO fills the HOMO–2 hole, thereby transferring the excess energy to the second electron in the HOMO which in turn is excited to the LUMO/+1. Our TDDFT calculations again confirm the orbital assignments and that this process is possible as $E(D_m^+) > E(D_1^+)$ (see Appendix E).

Similar to our above discussion of the anion, also the cationic molecule can now either remain charged and relax into the $D_0^+$ ground state via electron-hole recombination and $X^+$ emission, or it discharges into the $S_1$ state (dashed gray arrow (iv) in Fig. 4), from which it can radiatively decay to $S_0$ under $X^0$ emission. Interestingly, in case of the former process, the molecule can still discharge from $D_0^+$ ground state into $S_1$ and emit an additional $X^0$ photon, i.e., different from the scenario of the anionic molecule, the $X^+$ and $X^0$ emissions do not compete. This is again confirmed in the STML spectra shown in Fig. 1(b). At negative bias on 2 ML NaCl, there is strong $X^0$ emission but no $X^+$ emission. On 3 ML NaCl, the $X^0$ emission is about 3.5 times more intense, but also the $X^+$ emission emerges, with strong intensity. Furthermore, with increasing magnitude of the negative sample bias, both emissions increase in intensity (Fig. 2(a) and Fig. 7 in Appendix B). We note that the strongly tip-dependent non-integer power laws of the current-dependent emission intensities clearly indicate that all processes at $V_S < 0$ are plasmon-enhanced (Fig. 8 in Appendix B). This would also permit a plasmon-induced excitation of the cation from $D_0^+$ to $D_1^+$, following the previous interpretation that the $X^+$ emission is caused by a two-electron process [16]. While this could explain the up to $\alpha = 2$ values found for some tips, the fact that $X^+$ emission is only observed upon reaching a threshold voltage corresponding to the HOMO–2, also for these tips, leads us to conclude that plasmon-induced excitation from $D_0^+$ to $D_1^+$ does not play a role, presumably because the discharging process (iii) in Fig. 4 is much more efficient.

We note that our interpretation explains current STML results without considering the possibility of the triplet states, which are ignored in our many-body diagram. While excitation into triplet states may be possible, they do not contribute to the luminescence of ZnPc. First of all, the molecule has no strong spin-orbit coupling, hampering radiative decay from the triplet state. Secondly, triplet states are long-lived, typically on the order of μs. In comparison, typical tunneling rates are orders of magnitude faster. Hence, even if a molecule is excited to a triplet state, it will likely be charged by electron tunneling (and hence go into a doublet state) before radiative decay can occur. From this point of view, it remains unclear whether phosphorescence can be observed in STML experiments [34,35].

Finally, we would like to emphasize the general applicability of our proposed many-body energy diagram and the underlying interpretation that $X^+$ ($X^-$) emission from cationic (anionic) ZnPc (and likely other phthalocyanines) on ultrathin NaCl films is caused by resonant tunneling into higher-lying (un)occupied states followed by intramolecular electronic relaxations. For this, we refer to the STML results of ZnPc on NaCl/Au(111) from Doppagne *et al.* [16]. Compared to the NaCl/Ag(111) substrate of our study, the situation in the many-body diagram is reversed for the NaCl/Au(111) substrate: the HOMO and LUMO onsets there were found at about −0.9 V and +1.9 V, which would mean that the $D_0^+$ and $D_0^-$ states are below and above the $S_1$ state, respectively. The corresponding" many-body diagram would then then be a "flipped" version of that in Fig 4, predicting that both $X^0$ and $X^+$ emissions would only occur at a threshold voltage corresponding to tunneling out of the HOMO–2. Indeed, such a threshold was observed for $V_S < -2.2$ V on NaCl/Au(111), and STS data showed an increase of the differential conductance indicative of resonant tunneling from a lower-lying occupied MO. NaCl thickness-dependence is also suggestive of the competition between the $X^0$ and $X^+$ emissions. At positive bias, $X^0$ emission was reported at conditions where resonant tunneling into the LUMO occurs. There was no report on $X^-$ emission yet. However, based on our many-body diagram, we predict that it should be observable for larger positive voltages. Such an experiment may prove more difficult, as ZnPc is known to diffuse under such tunneling conditions. Therefore, the molecule may need to be stabilized as we did in this study, i.e., by either anchoring to a step edge [13] or by forming molecular chains [8,36].

## V. CONCLUSION

In summary, we observe bipolar electrofluorochromism of individual ZnPc molecules adsorbed on ultrathin NaCl films on Ag(111). Via the combination of STML and STS measurements, we demonstrate that the single-molecule fluorescence from cationic and anionic ZnPc can be rationalized by a combination of resonant tunneling, i.e. carrier injection into higher energy molecular orbitals, followed by Auger-like intramolecular relaxations. In conjunction with TDDFT calculations, we propose a many-body energy diagram that is able to describe all excitation and relaxation processes involved in the STML and that consistently explains all observations. We further show that this picture is generally expandable and applicable to other reported STML studies. The systematic approach presented here may serve as a blueprint for future studies, in order to identify excitation and relaxation pathways in single-molecule electroluminescence studies. Indeed, we note that a most recent study applied a similar approach on a different molecular emitter [37]. Our results emphasize that intermediate charged states may play an important role in the excitation and relaxation pathways of organic electroluminescent devices, beyond the double-barrier tunneling junction studied here. Also in OLEDs, electrons and holes are separately injected into the emission layer of the device, where fluorescent or phosphorescent emitter molecules are doped into a host matrix. Unless excitation of the emitter occurs via resonant energy transfer from surrounding host molecules [7,22,38], exciton formation involves an intermediate charged state, and the excitation and relaxation processes eventually leading to light emission may be more complex than merely involving the HOMO and LUMO. In this respect, it will be very interesting to expand future STML studies to substrates that are typical host materials in OLEDs.

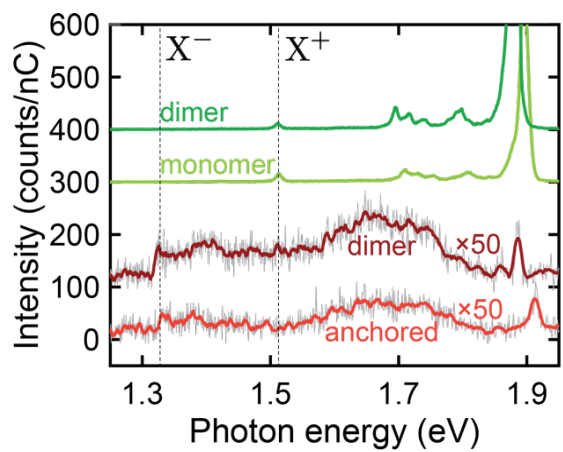

**Figure 5. STML spectra of monomer (anchored) vs. dimer ZnPc molecules on 3 ML NaCl/Ag(111).** STML spectra acquired with positive bias of anchored ZnPc monomer (red) and ZnPc dimer (dark red) ($V_S$ = +2.4 V, $I_t$ = 50 pA, acquisition time $t$ = 300 s), as well as acquired with negative bias (light and dark green, respectively ($V_S$ = −3.2 V, $I_t$ = 200 pA, acquisition time $t$ = 120 s). All spectra were acquired using the same tip and are offset for clarity. For the positive bias STML spectra, the raw data (gray) was 20-point Savitzky–Golay filtered.

**ACKNOWLEDGMENTS**

This work is part of the research program of the Foundation for Fundamental Research on Matter (FOM), which is financially supported by the Netherlands Organization for Scientific Research (NWO), as well as part of NWO project OCENW.M20.253. R.R. and N.L. acknowledge financial support from the Spanish State Research Agency grant (Project No. PID2021-127917NB-I00) funded by MCIN/AEI/10.13039/ 501100011033, as well as project ESiM 101046364 funded by the European Union. Views and opinions expressed are however those of the authors only and do not necessarily reflect those of the European Union. Neither the European Union nor the granting authority can be held responsible for them.

**APPENDICES**

**Appendix A: Evidence of direct charge injection to form $X^\pm$ emissions**

As evidence that in the dimer the $X^\pm$ emissions can only occur at the molecule where the tip is parked, we compare STML spectra acquired from monomers and dimers, shown in Fig. 5. We note that for this comparison, we used an anchored monomer in order to be able to apply large positive voltages without inducing diffusion. In the dimer, the $X^0$ emission is found to be redshifted compared to that of the monomer at both negative (green STML spectra) and positive (red STML spectra) sample bias. This implies that coherent excitation of both molecules occurs in the dimer due to dipole-dipole coupling [8,9,21]. Opposed to this, neither a redshift nor an intensity increase is observed for the $X^\pm$ emissions when comparing STML spectra of the dimer with those of the monomer. From this, we conclude that there is no detectable charge transfer between adjacent molecules in the dimer, hence excitons in a charged molecule (or trions) are localized to the molecule above which the tip is positioned [22].

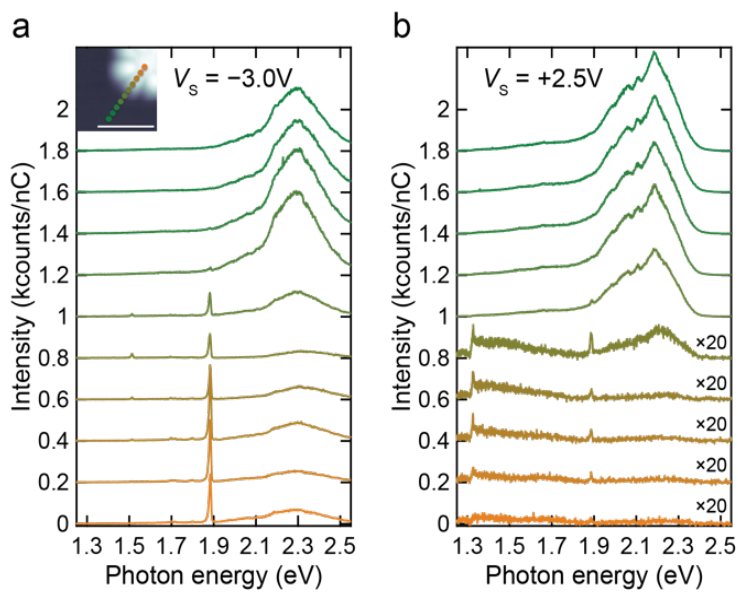

**Figure 6. Lateral distance-dependent STML spectra of ZnPc dimer on 3 ML NaCl/Ag(111).** (a) STML spectra acquired at $V_S = -3.0$ V. Lateral positions are color-coded and indicated by the dots in the inset constant-current STM image ($V_S = -2.5$ V, $I_t = 5$ pA, scale bar = 2 nm), the separation between each position being 250 pm. (b) STML spectra acquired at $V_S = +2.4$ V. All STML spectra were acquired in constant-current mode ($I_t = 100$ pA, acquisition time $t = 120$ s).

Figure 6(a) illustrates the lateral distance-dependent STML spectra of a ZnPc dimer adsorbed on 3 ML NaCl/Ag(111) acquired at $V_S = -3.0$ V. Only a broadband plasmonic resonance between about 1.9 and 2.5 eV can be observed when the tip is parked at a lateral distance of at least 1 nm away from the rim of the molecule. When the tip is parked closer to the molecule (ca. 0.5 nm away from the rim), a faint asymmetric Fano lineshape is observed at 1.9 eV (i.e. $X^0$ emission), which indicates that the molecule can be indirectly excited from $S_0$ to $S_1$ via plasmon-mediated energy transfer [13,15,16,39,40]. We note that this excitation channel should, in principle, be visible even at lateral distances < 1 nm from the rim of the molecule. However, in the experiments presented here the plasmon resonance was energetically not well aligned to the exciton energy. The $X^+$ emission only appears as soon as the tip is laterally positioned such that direct tunneling through the molecule occurs (as identified by the spatial extent in the STM topography).

A similar behavior is also found at $V_S$ = +2.5 V, as shown in Fig. 6(b). A broadband plasmonic resonance is found between 1.9 and 2.4 eV when the tip is parked on the 3 ML NaCl. The Fano feature at 1.9 eV appears when the tip is parked close to the rim of the molecule. The $X^-$ emission is only observed when direct tunneling into the molecule starts to occur. We conclude that the $X^\pm$ emissions can only be observed when the tip is parked on the molecule, and direct tunneling into/out of the MOs accessed at the respective threshold voltages is required to transiently charge the molecule [16].

**Appendix B: Emission thresholds and power law analysis for various tips**

In Fig. 7, we show the peak intensities of the $X^0$ and $X^\pm$ emissions from sample bias-dependent STML measurements carried out with four different tips (labelled B-E), which are different from the tip used for the data shown in the main text (tip A). The $X^0$ emission becomes visible at $V_S \lesssim -1.9$ V (see purple circle in Fig. 7(a)) but has a strongly increased intensity from $V_S \leq -2.2$ V. A similar enhancement of the intensity was also found for the emission of $H_2Pc$ and was attributed to resonant tunneling out of the HOMO [14,27]. We find that this resonant enhancement strongly varies with different tips, presumably due to variations in the plasmon enhancement of the light emission. The threshold of the $X^+$ emission (Fig. 7(b)) is also strongly tip-dependent, ranging between $-2.65$ V $\leq V_S \leq -2.90$ V. Overall, both $X^0$ and $X^+$ intensities increase with further decreasing $V_S$.

Peak intensities of $X^0$ and $X^-$ emissions for $V_S > 0$ are shown in Fig. 7(c) and (d), respectively. In both cases, emission thresholds are all found within a relatively narrow voltage range of $+2.05$ V $\geq V_S \geq +2.15$ V. Especially, for the $X^0$ peak we can rule out that emission starts at $V_S \gtrsim +1.9$ V, i.e., there is no noticeable plasmon-induced light emission at positive sample bias. We note that the thresholds slightly depend on the tunneling current $I_t$ used for the STML acquisition (see tip C in Fig. 8(c,d)), decreasing with increasing current. This can be explained in the picture of the double tunneling barrier, where with increasing current the tip-molecule separation (and hence the width of the vacuum barrier) is decreased, while the insulator barrier between the molecule and the metal substrate remains the same. As a consequence, the negative ion resonance (NIR), shifts toward the Fermi level [32,33]. In this respect, this observation is further indication that the thresholds for $V_S > 0$ are connected to resonant tunneling into a MO, which is required to observe both $X^0$ and $X^-$ emissions.

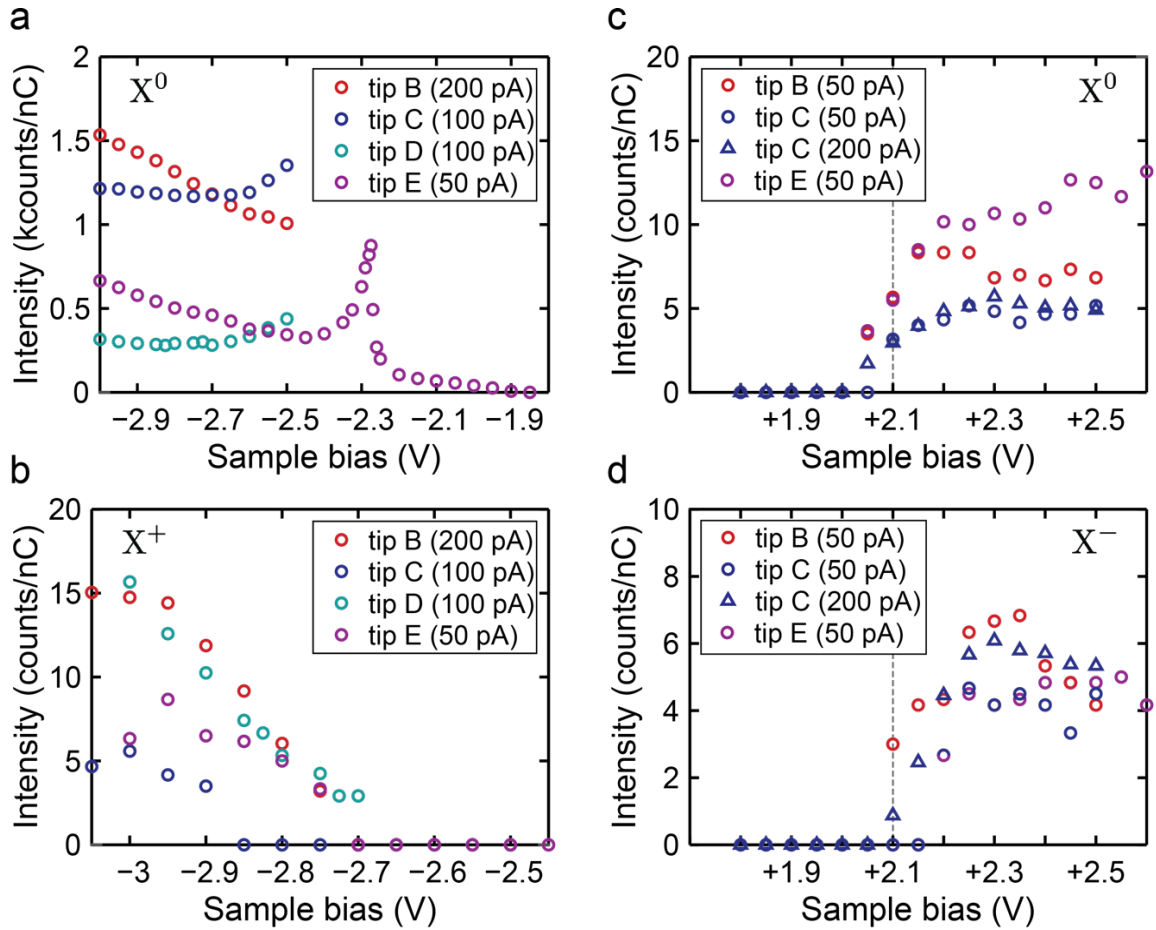

**Figure 7. Tip dependence of of $X^0$ and $X^\pm$ emission thresholds for ZnPc dimers on 3 ML NaCl/Ag(111).** (a) Peak intensity of the $X^0$ emission vs. sample bias for $V_S < 0$. For all tips, it becomes visible at $V_S \lesssim -1.9$ V but has a strongly increased intensity from $V_S \leq -2.20$ V that depends on the tip. (b) Peak intensity of the $X^+$ emission for $V_S < 0$. The observed threshold varies with different tips between −2.65 V and −2.90 V. (c) Peak intensity of the $X^0$ emission and (d) of the $X^-$ emission for $V_S > 0$. In both cases, a almost tip-independent onset of the threshold is observed between +2.05 V and +2.15 V. All measurements were acquired in constant-current mode using stabilization currents as stated in the legends.

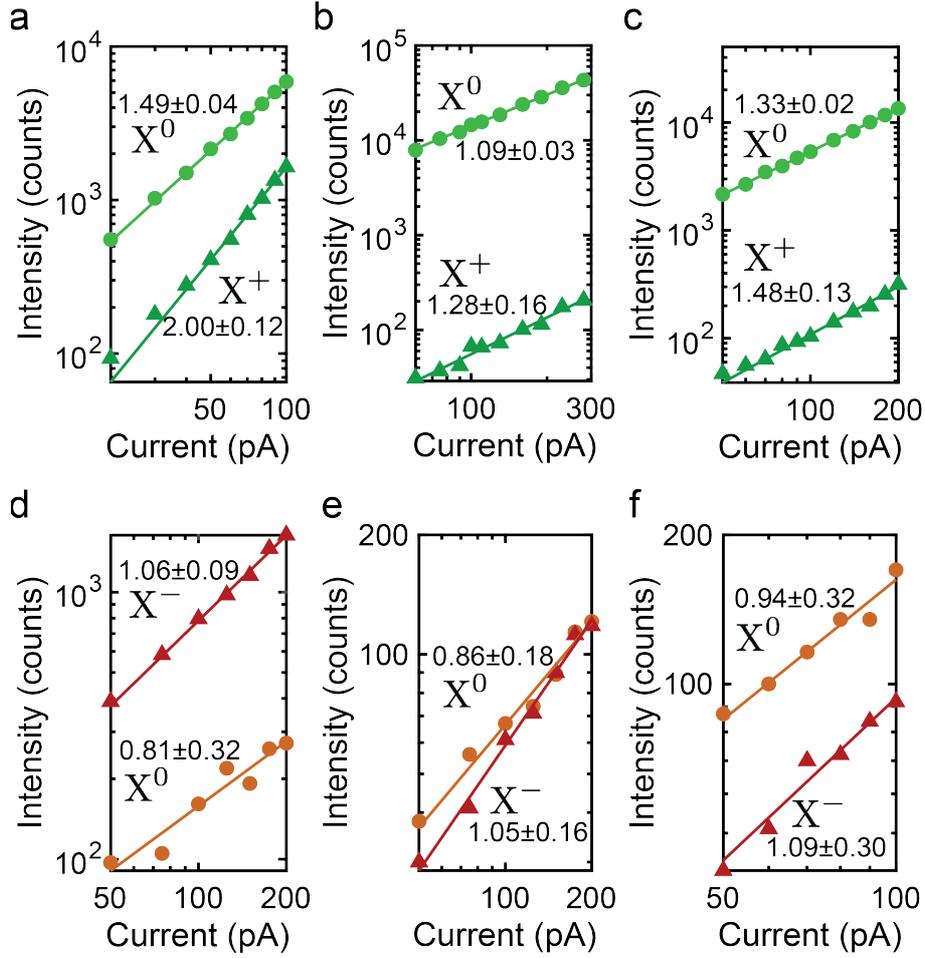

**Figure 8. Power law analysis of ZnPc.** (a-c) Log-log graph of current-dependent intensity of $X^0$ and $X^+$. (d-f) Log-log graph of current-dependent intensity of $X^0$ and $X^-$. The sample biases are (a) $V_S$ = −3.2 V, (b) $V_S$ = −3.0 V, (c) $V_S$ = −3.0 V, (d) $V_S$ = +2.5 V, (e) $V_S$ = +2.5 V, and (f) $V_S$ = +2.4 V. The intensities in (a) and (d) were acquired from the spectra by integrating the total photon counts in the range ±2 meV around the peak position. The intensities shown in (b-c) and (e-f) were acquired from the spectra at the maximum of the peak position.

We also analyzed the tip dependence of the current-dependent STML intensity [16,18,26,27]. Fig. 8(a-c) shows $\Im_{photon}$ vs. $I_t$ in a log-log graph, for both $X^0$ (green circle) and $X^+$ (green triangle), acquired for various tips at (a) $V_S$ = −3.2 V, (b) $V_S$ = −3.0 V, and (c) $V_S$ = −3.0 V. By fitting of the data with the power law $\Im_{\text{photon}} \propto I_t^\alpha$, we obtain $\alpha(X^0)$ and $\alpha(X^+)$ values shown in Fig. 8(a-c). We find that the exponents for both $X^0$ and $X^+$ are strongly tip-dependent, varying in the range of $1.09 \leq \alpha(X^0) \leq 1.49$ and $1.28 \leq \alpha(X^+) \leq 2.00$, respectively. The non-integer value can be interpreted as a plasmon-enhanced single-electron process [27], in accordance with previous interpretations [13,14]. Regarding $X^+$ in Fig. 8(a), the power of 2 suggests a two-electron process. In fact, this has been suggested previously [16,17]. However, the highly important role of signal enhancement by nanocavity plasmons (NCPs) already shown above may also play a role in the power law behavior of both $X^0$ and $X^+$ emissions at negative bias [21,28-30]. Indeed, $\alpha$ varies dramatically for different tips (exhibiting different NCP resonances), but on average clearly is below a value of 2. We note that for same tips but different sample biases, $\alpha$ also changes, further emphasizing the role of the NCPs in STML at negative sample bias.

On the other hand, at positive sample bias the power law analysis gives a very clear result. Fig. 8(d-f) illustrates the tip-dependent measurements of $X^0$ (orange circle) and $X^-$ (red triangle), acquired at (d) $V_S$ = +2.5 V, (e) $V_S$ = +2.5 V, and (f) $V_S$ = +2.4 V along with the fitting results (solid lines). The values of $\alpha(X^0)$ and $\alpha(X^-)$ are shown in Fig. 8(d-f). The larger errors are due to the much smaller STML yields. The results show that the exponents barely vary with different tips or different $V_S$, and in fact are all within the error bars. On average, we find $\alpha(X^0) = 0.87 \pm 0.16$ and $\alpha(X^-) = 1.07 \pm 0.12$. These values strongly suggest that both emissions are caused by one-electron processes.

## Appendix C: $X^0$ and $X^-$ threshold vs. 2nd NIR onset of dimer on 2 ML NaCl/Ag(111)

To connect the threshold voltage for STML to the MO onset energy at positive sample bias, we also perform bias-dependent STML and STS measurements on the ZnPc dimer adsorbed on 2 ML NaCl/Ag(111), shown in Fig. 1(a). Here, we utilize the fact that on 2 ML vs. 3 ML the molecular ion resonances are slightly shifted closer to the Fermi energy on the thinner NaCl film due to increased electronic polarization by the metal substrate [31]. Indeed, we find that the onset of the first NIR (i.e. the LUMO) shifts from $V_S$ ≈ +0.75 V on 3 ML to $V_S$ ≈ +0.65 V on 2 ML NaCl(Ag(111). In Fig. 9, the differential conductance around the onset of the second NIR of ZnPc on 2 ML NaCl is shown as a blue curve. Its onset is found at $V_S$ ≈ +2.05 V, i.e. also about 0.1 V below that found on 3 ML NaCl. The red circles and triangles show the STML peak intensity vs. $V_S$ of $X^0$ and $X^-$, respectively. Their thresholds are also both shifted to the bias voltage of $V_S$ ≈ +2.00 V.

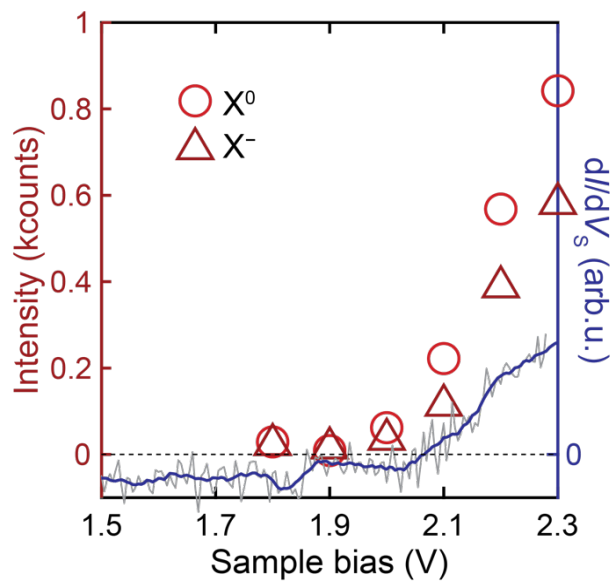

**Figure 9. Differential conductance spectrum vs. the thresholds of $X^0$ and $X^-$ emission of ZnPc dimer adsorbed on 2 ML NaCl/Ag(111).** The threshold of $X^0$ (red circle) and $X^-$ (red triangle) is found at sample bias $V_S$ = +2.1 V. The onset of the second NIR is located around $V_S$ = +2.1 V (blue curve). STML spectra were acquired with different sample bias (taken in constant-current mode, tip position is indicated in Fig. 1(a), $I_t$ = 100 pA, $t$ = 120 s). The intensities are extracted from the spectra by integrating the total photon counts in the range ± 2 meV around the peak position. For the STS, the feedback loop was opened at the center of molecule at $I_t$ = 100 pA, and sample bias $V_S$ = +2.4 V.

**Appendix D: Theoretical details and molecular orbital assignment**

In order to get more insight into the adsorbed molecule and its electronic structure we performed a series of DFT calculations for a single ZnPc molecule on the surface. We performed the calculations with the VASP code [41], using the PAW method [42] and planewave basis sets with an energy cutoff of 400 eV. We applied the PBE exchange and correlation functional [43] complemented with the D3 method [44] to account for the missing van der Waals interactions. For the simulation of the NaCl/Ag(111) system we used three NaCl layers in a (5x6) unit cell, with four Ag layers to simulate the Ag(111) surface (Fig. 10(a) and (b)). We employed the calculated NaCl lattice constant (5.65 Å) and adjusted the Ag lattice constant accordingly, resulting in a lattice constant for Ag of 4.00 Å (a 2% reduction compared to the experimental one). We relaxed all the atoms except the two bottom Ag layers until all the forces were smaller than 0.01 eV/Å. For the STM simulations we applied the Tersoff-Hamann approximation [45] using the method by Bocquet *et al.* [46] implemented in STMpw [47].

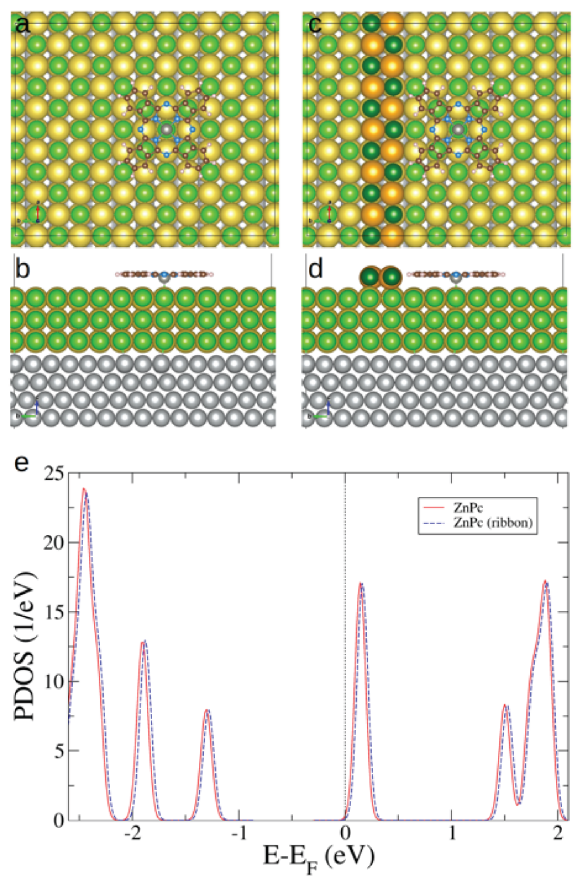

**Figure 10. Unit cells and PDOS for DFT calculations of ZnPc on NaCl/Ag(111).** Top (a) and side (b) views of the unit cell used in the calculation of a ZnPc molecule on NaCl/Ag(111). In (c) and (d) the same is shown for a ZnPc molecule anchored to a ribbon of NaCl atoms. (e) Projected density of states for ZnPc with and without a ribbon.

We placed one ZnPc molecule on NaCl/Ag(111). Comparing the total energies of different adsorption sites, we determined that the most stable position corresponded to the Zn atom on top of a Cl atom, with a rotation of 10º with respect to the molecule fully aligned with the NaCl axes. This angle agrees with the experimental observations, where the molecule rapidly shuttles between ±11º when it is not anchored. We have also simulated the experimental situation where the molecule is anchored to a step edge. We have used the geometry shown in Fig. 10(c) and (d), where the step edge is simulated by an additional ribbon of NaCl atoms on top of the 3 ML NaCl. The ZnPc molecule is placed next to it with the Zn atom on top of a Cl atom. In order to check if the anchoring changes the electronic properties of the molecule, we have compared the projected density of states (PDOS) of ZnPc on the same adsorption position with and without the ribbon. As seen in Fig. 10(e), the PDOS of both situations is identical except for a small shift.

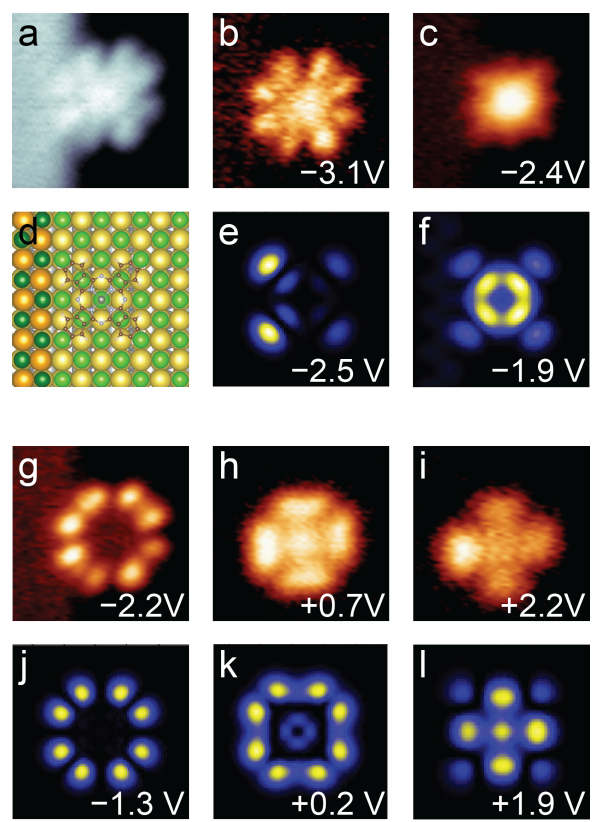

**Figure 11. Molecular orbital mapping and DFT calculation of an anchored ZnPc adsorbed on 2 ML NaCl/Ag(111).** (a) Constant-current STM image of ZnPc molecule anchored against to 3 ML NaCl step edge ($V_S$ = −2.4 V, $I_t$ = 10 pA). (b-c,g-i) Constant-height differential conductance maps. Sample biases are indicated in the figures. The feedback loop was opened at set current $I_t$ = 100 pA and sample bias (b) $V_S$ = −3.1 V, (c,g) $V_S$ = −2.4 V, (h) $V_S$ = +0.9 V, and (i) $V_S$ = −3.1 V above the center of the molecule. (d) Stick-ball model of the anchored molecule. (e-f,j-l) DFT simulated constant-height differential conductance maps at voltages as stated.

To complement the analysis of the onset voltages of MOs within a dimer, as done in the main text, we also show the same analysis for a monomer, and we compare differential conductance maps to DFT calculations, as shown in Fig. 11. Here, an ZnPc molecule adsorbed on 2 ML NaCl/Ag(111) and anchored to a 3-ML step edge is used in the experiment (Fig. 7(a)) because the isolated monomer shows shuttling motion which makes it difficult to identify the MOs [8]. We compare with the DFT results from a molecule adsorbed on 3 ML NaCl next to a 4-ML NaCl stripe (d). Experimental results of ZnPc on 3 ML NaCl anchored to a 4-ML step edge are comparable, leading only to small differences of peak positions in d$I$/d$V$ spectra up to 0.1 V, i.e. smaller than the peak widths (see also Appendix C). We further note that the step edge leads to slight asymmetries in the differential conductance of the orbital maps, seen both in experiments and calculations. We verified that the step edge does not impact the qualitative orbital shapes and order by also calculating ZnPc on the free 3-ML terrace as well as by comparison with dimer maps (Fig. 3) [13].

At positive bias, the map of the first NIR (Fig. 11(h)) is characterized as the degenerate LUMO and LUMO+1 (k). The map of the second NIR (i) shows a more cross-like distribution. By comparing to the DFT calculations, this may be attributed to the LUMO+3 (l). A map corresponding to the calculated LUMO+2 (not shown) was not found in the experiments, but we note that in the calculations it looks rather similar to and hence may be difficult to distinguish from the degenerate LUMO/+1. We cannot rule out that in the experiments, the LUMO+2 may be hidden in the NDR region (see below). In this context, the STS data in Fig. 3(a) shows a shoulder at the trailing edge of the NIR peak above $V_S$ = 1 V, which one may be tempted to identify as another MO. However. Miwa *et al.* showed for H$_2$Pc that this can readily be explained by vibronic excitations [14]. Due to this unclear situation, we refer to this second NIR as the LUMO+$x$.

At negative bias, according to the spatial maps, the first (Fig. 11(g)) and the second PIR (c) can be assigned to the HOMO (j) and HOMO−1 (f), respectively, the latter being mainly located at the central Zn atom and exhibiting significant *d*-orbital character. At even lower sample bias, at least the third PIR is found, but not exhibiting a sharp onset energy (see Fig. 3(a)). A spatial map at $V_S$ = –3.1 V is shown in Fig. 11(b). It can be attributed to either the HOMO–2 (e) or even a lower lying MO. Indeed, the STS spectrum shows a much broader resonance below –3 V, which could be due to the presence of two orbitals that strongly overlap, owing to the electron-phonon broadening effect [14]. Therefore, we refer to the third PIR as the HOMO–$x$.

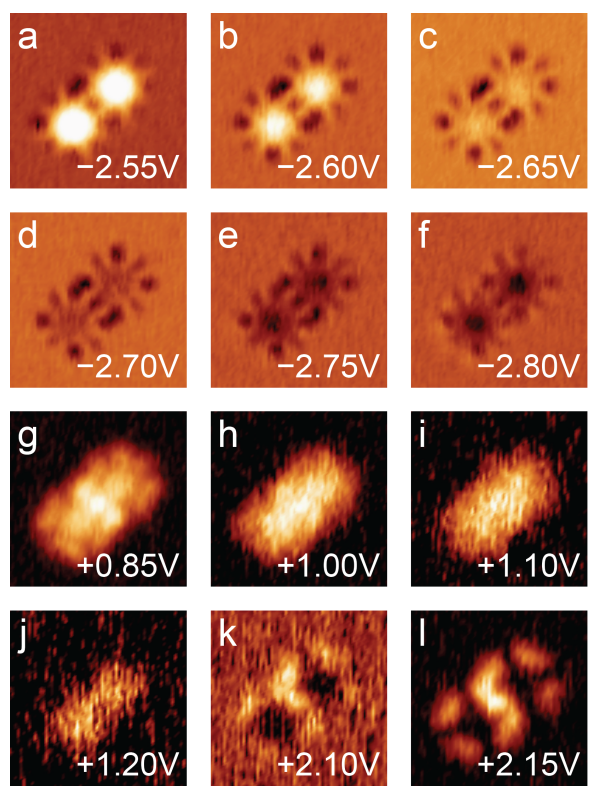

**Figure 12. Molecular orbital mapping with increment bias steps.** Constant-height differential conductance maps at sample biases as indicated. Feedback loop was opened at: (a-f) $V_S$ = −2.8 V, $I_t$ = 50 pA; (g-j) $V_S$ = +1.2 V, $I_t$ = 50 pA; (k-l) $V_S$ = +2.3 V, $I_t$ = 50 pA.

To clearly see a transition from one MO character to the next as the voltage is changed, we also carried out the spatial mapping of the ZnPc dimer with smaller increments of $V_S$ in both NDR regions seen in Fig. 3(a). For negative sample bias (Fig. 12(a-f)), we indeed find a transition from the HOMO−1 to another MO as we decrease $V_S$ within the bias range of the $X^+$ threshold (−2.65 V ≤ $V_S$ ≤ −2.90 V). The latter is similar to the map shown in Fig. 11(b) but with reversed contrast due to the NDR. For positive sample bias, Fig. 12(g-j) shows maps at voltages where the first NIR is observed in STS spectra. With increasing voltage, the maps become blurrier. This may be caused by electron-phonon coupling leading to a broad Gaussian line shape toward higher energies [14,31,32], and provides further indication that we cannot identify another MO in the range where we observed a shoulder in STS (Fig. 3(a)). Above 2 V, the emergence of another MO is again observed in the NDR region (Fig. 12(k,l)) at the onset of the second NIR.

**Appendix E: Time-dependent density functional theory results**

Using the Casida method [48] as implemented in VASP [41,42] we can calculate some of the molecular excitation energies for different charge states. Here, the ground state on the anion corresponds to an extra electron occupying the LUMO, and the ground state of the cation has one electron less in the HOMO. The energy diagram of spin-dependent orbital energies of anionic and cationic ZnPc in the ground state are shown in Fig. 13. The Casida method allows for a time-dependent density functional theory (TDDFT) evaluation of the excited states that correspond to a single electron-hole excitation in the molecule (see double arrows in Fig. 13). We note that our TDDFT calculations did not aim at a quantitative but a qualitative analysis, in order to rationalize the experimental results. Besides, deviations between TDDFT and experimental results in the range of ~0.2-0.4 eV are common [14,49]. The anion excitation energies start at 1.117 eV, i.e. much lower than the optical spectra. To identify which excitations can lead to light emission, we also computed their respective oscillator strengths. For the anion, the excitations at 1.117 eV and 2.165 eV have zero oscillator strength. The first non-zero oscillator strength takes place for the third excitation at 2.221 eV.

We first look at different excitations of the anionic doublet state $D^-$. When the resonant electron tunneling into the second NIR occurs, the molecule is excited into a higher anionic doublet state $D_n^-$. From TDDFT, we identify this as the state with excitation energy $\Delta E(D_n^-)$ = 2.389 eV (Fig. 13). Further analysis of the TDDFT results reveal that this excited state can be described by an electron in the LUMO+3 (or higher) and a hole in the LUMO, as shown in the simplified diagram of the $D_n^-$ state in Fig. 4. Due to the uncertainty of the correct assignment of this orbital, we refer to it as the LUMO+*x* (where *x* ≥ 2). Next, in a single-particle picture, the electron relaxes from the LUMO+*x* to one of the LUMO/+1 levels, and the energy difference allows for another electron in the HOMO to be excited to one of the LUMO/+1 levels. This Auger-like intramolecular relaxation process leads to a double occupation of the LUMO/+1, while leaving a hole behind in the HOMO, as shown in the simplified diagram of the $D_1^-$ state in Fig. 4. Comparing with TDDFT, we indeed find that the 2.221 eV excitation in Fig. 13 assigns an additional electron to the LUMO and a hole in the HOMO. In other words, the total energy of this doublet state decreases to $\Delta E(D_1^-)$ = 2.221 eV, i.e. $\Delta E(D_n^-) > \Delta E(D_1^-)$. Alternatively, the excitation at 2.255 eV places an electron in the LUMO+1 and a hole in the HOMO, which also has non-zero oscillator strength, but we note that this excitation is very close in energy to the $D_1^-$ state. This means that the suggested intramolecular relaxation process shown in Fig. 4 is energetically allowed. Once in the $D_1^-$ state, the molecule can remain charged and relaxes into the $D_0^-$ ground state, the excess energy is emitted as a $X^-$ photon.

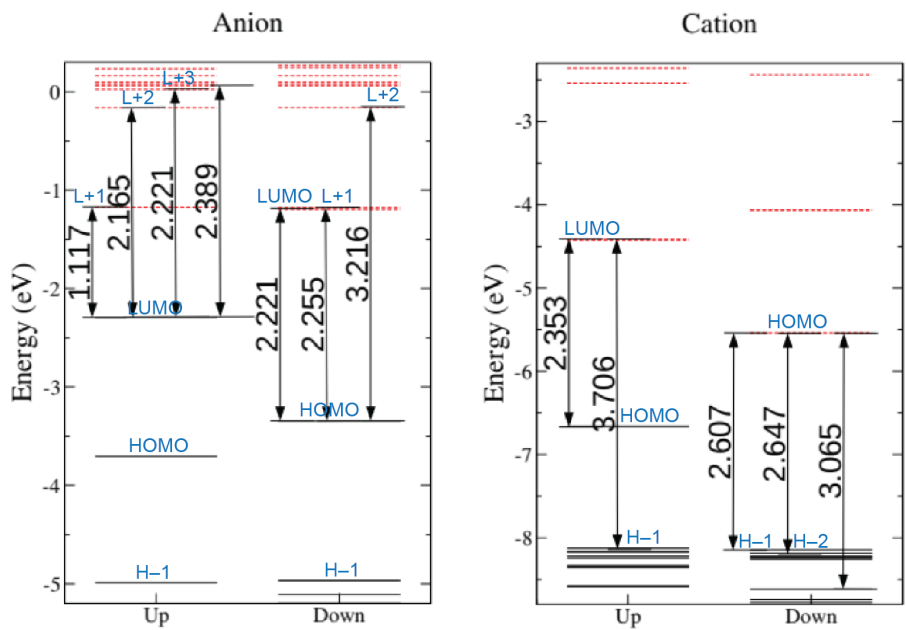

**Figure 13. TDDFT calculation results**. The plots show spin-dependent orbital energies of anionic and cationic ZnPc in the ground state. Black solid lines refer to occupied states, red dashed lines to empty states. Orbital labels (blue) are with respect to the neutral molecule. TDDFT-calculated excitation energies for the lower-energy transitions are also given, and the double-arrows show in which orbitals the electron-hole pair is (mainly) located. For the anion, the first excitation with non-zero oscillator strength is that at 2.221 eV; for the cation, it is the excitation at 2.353 eV.

Next, we evaluate the excitation energies of the cation doublet state $D^+$. When the resonant electron tunneling out of the second (or higher) PIR (HOMO−x) takes place, the molecule is excited into a high cationic doublet state $D_m^+$. From TDDFT, we identify this as the state with excitation energy 2.647 eV (Fig. 13), and it can be assigned mainly to a hole in the HOMO–2 and an electron in the HOMO, as shown in the simplified diagram of the $D_m^+$ state in Fig. 4. In a single-particle picture, the hole in the HOMO–2 is then filled by an electron from the HOMO level, accompanied by an Auger-like intramolecular relaxation process in which the other HOMO electron is excited to one of the LUMO/+1 levels, as shown in the simplified diagram of the $D_1^+$ state in Fig. 4. According to TDDFT, the excitation energy to this doublet state decreases to $\Delta E(D_1^+) = 2.353$ eV $< \Delta E(D_m^+)$. Also here, the relaxation process is hence energetically possible. Once in the $D_1^+$ state, the molecule can remain charged and relax into the $D_0^+$ ground state, emitting the excess energy as a $X^+$ photon.